\documentclass[sigconf]{acmart}
\usepackage{booktabs} 
\usepackage{tabu}
\usepackage{placeins}
\usepackage{subfig}
\usepackage{graphicx}
\usepackage{graphics}
\usepackage{enumitem}
\usepackage{balance}

\begin{document}
\title{Illuminating an Ecosystem of Partisan Websites}

\author{Shweta Bhatt}
\affiliation{%
  \institution{King's College London}
  \streetaddress{}
  \city{London} 
  \state{UK} 
  \postcode{}
}
\email{shweta.bhatt@kcl.ac.uk}

\author{Sagar Joglekar}
\affiliation{%
  \institution{King's College, London}
  \streetaddress{}
  \city{London} 
  \country{UK}}
\email{sagar.joglekar@kcl.ac.uk}

\author{Shehar Bano}
\affiliation{%
  \institution{University College London}
  \city{London}
  \country{UK}}
\email{s.bano@ucl.ac.uk}

\author{Nishanth Sastry}
\affiliation{%
  \institution{King's College, London}
  \streetaddress{}
  \city{London} 
  \state{UK} 
  \postcode{}
}
\email{nishanth.sastry@kcl.ac.uk}


\begin{abstract}
This paper aims to shed light on alternative news media ecosystems that are believed to have influenced opinions and beliefs by false and/or        biased news reporting during the 2016 US Presidential Elections. We examine a large, professionally curated list of 668 hyper-partisan websites and their corresponding Facebook pages, and identify key characteristics that mediate the traffic flow within this ecosystem. We uncover a pattern of new websites being established in the run up to the elections, and abandoned after. Such websites form an ecosystem, creating links from one website to another, and by `liking' each others' Facebook pages. These practices are highly effective in directing user traffic internally within the ecosystem in a highly partisan manner, with right-leaning sites linking to and liking other right-leaning sites and similarly left-leaning sites linking to other sites on the left, thus forming a filter bubble amongst news \emph{producers} similar to the filter bubble which has been widely observed among \emph{consumers} of partisan news. Whereas there is activity along both left- and right-leaning sites, right-leaning sites are more evolved, accounting for a disproportionate number of abandoned websites and partisan internal links. We also examine demographic characteristics of consumers of hyper-partisan news and find that some of the more populous demographic groups in the US tend to be consumers of more right-leaning sites.



\end{abstract}

%
%


\copyrightyear{2018}
\acmYear{2018}
\setcopyright{acmcopyright}
\setcopyright{iw3c2w3}
\acmConference[WWW '18 Companion]{The 2018 Web Conference Companion}{April
23--27, 2018}{Lyon, France}
\acmBooktitle{WWW '18 Companion: The 2018 Web Conference Companion, April
23--27, 2018, Lyon, France}
\acmPrice{}
\acmDOI{10.1145/3184558.3188725}
\acmISBN{978-1-4503-5640-4/18/04}


\newcommand\mypara[1]{\textbf{\textit{#1}}.}

\maketitle
\section{Introduction}


The rise of the Internet and social media platforms has given a new dimension to the democratic process. From a political science perspective, the idea of open participation platforms like blogs and social media websites have been shown to have high impact in the democratic process~\cite{lawrence2010self} and may well be instrumental in shaping the elections of the future, as demonstrated by the 2016 US elections.
As such, it is crucial for political and social scientists to assess the modalities and attitudes towards partisan media, and to understand their modus-operandi. 

So far, research has focused on evaluating the role of \emph{social} media and platforms like \texttt{Reddit} and \texttt{4chan} in the process of sharing mainstream and alternative news on the web
and the lexical value of the language used by fake-news content~\cite{zannettou2017web,nguyen2012sources,horne2017just,finn2014investigating,kwon2013prominent,starbird2017examining}. This research has revealed a deep divide between views of the so-called `right' or `conservatives', and the `left' or `liberals'. 

This paper attempts to shed first light on hyper-partisan \emph{news} websites that have arisen in parallel to this divided social media discourse. `Hyper-partisan' is a moniker that is believed to originate from a recent article in the New York Times Magazine~\cite{nytmag} and refers to reporting that departs from traditional notions of journalistic balance, and presents a biased picture of one side of a political debate. In the context of the 2016 US Elections, overt support  of the Democratic (respectively Republican)  party or their candidate(s) is taken to represent   a liberal or left-leaning  (resp.\ a conservative, or right-leaning) bias. We are interested in understanding the producers of hyper-partisan news, the consumers who visit them, and the dynamics of how Web traffic is directed towards such sites.

We base our study on an ecosystem of 668 hyper-partisan\footnote{In the rest of this paper, we interchangeably use the words \emph{partisan} and \emph{hyper-partisan} in connection with websites in this dataset. Operationally, the judgement of whether a news site is `partisan,' and if so, whether it is right- or left-leaning, was made by journalists at Buzzfeed News as part of a careful and impartial manual coding process. This is standard social science practice. Some results in this paper were first reported as part of our contributions to the  Buzzfeed News article \cite{criagFakenews}.} websites and associated Facebook pages shared with us by Buzzfeed News as part of a recent collaborative investigation connected with the 2016 US Elections. Building on this dataset\footnote{The dataset we collected is made available at \texttt{http://bit.ly/partisan-news-data} for non-commercial research usage.}, we collect data from \texttt{Alexa.com} about the size of the traffic to these sites, from Facebook about the structure of connections (likes, etc.) between the different Facebook pages, and from URL shorteners like \texttt{bit.ly} about the clickthrough rates for short URLs associated with news stories. Our central approach is to make use of the reliable and impartial manual coding of websites into `left' and `right' by our Buzzfeed collaborators, and use our augmented data to shed light on partisan behaviour on both sides of the political spectrum. 




Our first main finding is that there is a large amount of linking between partisan sites that refer traffic between each other. These referrals mostly stick to one side of the political debate -- right-leaning websites refer traffic to other right-leaning sites;  left-leaning sites link to left-leaning ones. In other words, among these \emph{producers} of hyper-partisan news, we find  echo chambers similar to those which have been widely discussed (e.g., in~\cite{garimella2017reducing}) among \emph{consumers} on social media. Unsurprisingly, we find a similar right-left cliquishness when we study the corresponding Facebook pages of these news sites, and ask which Page is `liked' by which other page. Completing the picture, the audiences of the Facebook pages are similarly divided across right-left lines, which is closer to the traditional notion of echo chambers in social media. 

Next, examining domain ownership records, we find evidence of a significant boom and decline of hyper-partisan news corresponding with  the 2016 US Election cycle -- a remarkable number (nearly one third) of new sites, especially right-leaning ones, were registered in 2016, during the run-up to the elections. Several are from Macedonia, providing independent confirmation of the role of Macedonian teenagers in propagating false and biased information~\cite{wired}. Similarly, a disproportionate number of these sites also ``die out'' and lose popularity in Alexa Traffic Rankings in the period between the Nov 2016 elections and President Trump's inauguration. 

This establishment and abandonment of website domains is  stronger amongst right-leaning sites, suggesting that the American Right or Conservative populace was more strongly influenced by hyper-partisan sites during the election. Complementary to this, we examine the demographics of the audience of hyper-partisan sites and find  that Americans who visit these hyper- partisan sites are more right-leaning: Sampled mean demographic audience sizes for some of the most populous demographic groups of USA (e.g., white caucasians, or middle income ranges) are higher for right-leaning sites. 

Collectively, these results shed first light on an ecosystem of websites that affected the 2016 US Elections by presenting a partisan viewpoint. The rest of this paper is structured as follows.  \S\ref{sec:Related} places our paper in the context of related work. \S\ref{sec:Dataset} describes the dataset provided to us by Buzzfeed and our additional data collection efforts with Alexa, bit.ly, and Facebook. \S\ref{sec:Owners} examines the websites studying the characteristics of its owners and the evolution of traffic to these sites. \S\ref{sec:Alexa} presents a complementary picture, studying the demographics of the \emph{audience} of these sites. \S\ref{sec:TrafficFlows} examines the producer side filter-bubble caused by the partisan referral of traffic from one website to another of the same political leaning. \S\ref{sec:Facebook} studies demographics and traffic flows from the perspective of Facebook pages associated with these sites, finding similar evidence of internal referrals and increased activity on the right. \S\ref{sec:Conclusion} concludes.

\section{Related Work}
\label{sec:Related}
We contextualize our work by providing a brief discussion of three key themes of research in this area. We survey the previous works which looked at \textit{Role of News sources}, \textit{Role of echo chambers and filter bubbles} and \textit{Role of social media in dissemination of fake news}. To this extent  our work augments and extends previous literature by presenting a data-driven analysis of the producers and consumers of partisan websites, and the traffic forwarding behaviour that underpins their sustainability and supports on-line echo chambers. 

\mypara{Role of news sources}
Classically political and social science heavily relies on survey-based inferences.
There have been important studies on partisan news media in the US over the past few years. These methods help the community get important insights into the divides in audience attitudes towards media \cite{pewSurveyNews3} and the role of social media in news dissemination \cite{pewSurveyNews1}. However, what they lack is an understanding of how news websites are shaping these attitudes and trends, and what role social networks play in facilitating the spread of partisan information.  
A recent data-driven work~\cite{Allcot} provides insights into the role of fake news in US elections. The authors use interesting sources of data, such as a crowd-driven website \texttt{Snopes.com} that refutes fake rumors and  conspiracy theories. They further find predictors for believability on face news. Their study is highly selective towards the fake news phenomenon, and ignores the effect  of online news sources (such as  \textit{Breitbart}, \textit{Infowars}, \textit{ThinkProgress}, \textit{OccupyDemocrats}) that promote partisan content which may not necessarily be fake.
We find in our study that these sources have big enough presence to shape traffic and grab click-through rates as high as 4\% on their posts. Karamshuk \textsl{et al.}~\cite{partisanDmytro} showed quantitative evidence for biased reporting in news and social media across national divides in the context of the recent conflict between Russia and Ukraine. Our work in contrast applies to ideological divides within a nation, and during a national election rather than a conflict situation.

\mypara{Role of echo chambers and filter bubbles}
Echo chambers refer to the phenomenon when people do not subscribe to views that do not agree with their own. Coupled with recommendation system algorithms, a self-reinforcing ecosystem emerges where users get fed information that suits their opinions~\cite{flaxman2016filter}. 
Pariser~\cite{pariser2011filter} argues that the ubiquitous nature of technology that personalizes every search and every interaction has created this ``In the Loop'' process.
This has inspired a series of works on recommendation systems for Facebook users that have revealed both the presence of filter bubbles and awareness among some users about their existence~\cite{rader2015understanding}. 
Recent work for probable remedies ~\cite{garimella2017reducing} explores breaking the polarized echo chambers by exposing users to opposing sides of discourse. Our work can help such fledgling efforts by helping shed light on the nature of the divide between different viewpoints.

\mypara{Role of social media in dissemination of fake news}
The issue of fake news and conspiracies has received considerable attention since the 2016 US Presidential election. 
Various studies~\cite{zannettou2017web,nguyen2012sources,horne2017just,finn2014investigating,kwon2013prominent} have explored the role of social media and platforms like Reddit and 4chan in the process of sharing mainstream and alternative news on the web---but these focus on the lexical value of the language used by fake-news content. 
Starbird et. al~\cite{starbird2017examining} explore the alternative media domain networks by analyzing 58M total tweets related to mass shooting events for a period of about 10 months beginning in January 2016, and provide a qualitative analysis of how alternative news sites propagate and promote false narratives while mainstream media refrains from such behaviour. They show the presence of domain clusters that control the flow of information with the alternative news ecosystem and their political polarization. As the scope of this work is limited to the Twitter activity of influential alternative news websites, it does not capture the referer traffic patterns between social network pages or the actual web pages. Our work complements the findings of this paper using other modalities of data and other datasets. 
\section{Dataset Description}
\label{sec:Dataset}

Our primary dataset comprises of a curated list of partisan websites and associated Facebook pages, augmented with traffic and audience engagement information using  third party websites and APIs like Alexa\footnote{https://www.alexa.com/about} and Facebook Graph API\footnote{https://developers.facebook.com/docs/graph-api/}. 
\subsection{Buzzfeed Dataset} 
As mentioned previously, Buzzfeed News provided us with a dataset which was collected and curated by them in the context of an investigation into partisan news in the 2016 US Presidential Elections. This dataset consists of two main parts: a list of 668 websites which were classified by them as partisan, and a list of Facebook pages which are associated with these websites.

\subsubsection{Partisan websites dataset} 
The Buzzfeed data set contains of a list of $668$ unique partisan sites, of which $489$ were classified as left-leaning and $179$ as right-leaning after a careful manual examination by two journalists from Buzzfeed News. These 668 websites form the basis of the rest of our study, and we take as ground truth the journalists' professionally informed categorisation of the websites into `left' and `right'. In addition, the following key attributes were collected about  each site: registration date, Facebook Id,  owner firstname, lastname and company, political category, Google analytics and Google Adsense codes 

\subsubsection{Facebook}
Facebook has become an important platform for communications. Many news organisations, including virtually all major mainstream newspapers maintain associated Facebook pages. The  Facebook page for each website in our dataset was identified using either direct links from the website itself, or using the CrowdTangle browser plugin to identify Facebook pages which share all the content from the corresponding website. 507 of the 668 websites have public Facebook pages that could be programmatically accessed. For each accessible Facebook page, the dataset contains engagement information of each post--- status id, status message, status link, status type, timestamp when the post was published, no. of reactions, no. of comments and no. of shares of 4M posts between January 1st, 2015 and March 31st, 2017. 
 Table~\ref{tab:fb_data_overview} provides an overview of the Buzzfeed Facebook dataset.

\begin{table}[!htb]
\begin{center}
\footnotesize
\begin{tabu} to 0.35\textwidth {  X[l] | X[r] | X[r]  }
 \toprule
 \textbf{Metric} & \textbf{Total} & \textbf{Average} \\
 \midrule
 \# Fans & 326M & 0.67M \\
 \# Posts & 4.1M & 8514 \\
 \# Post Reactions & 0.59M & 1163 \\
 \# Post Comments & 66924 & 132 \\
 \# Post Shares & 0.3M & 605 \\
 \bottomrule
\end{tabu}
\caption{Buzzfeed Facebook Dataset Overview}
\label{tab:fb_data_overview}
  \vspace{-0.8cm}
  \end{center}
\end{table}

\subsubsection{On the Quality and Completeness of the Buzzfeed Data}
To our knowledge, this dataset represents one of the most comprehensive lists of partisan news websites associated with the 2016 US elections. Buzzfeed's article~\cite{criagFakenews} publishes full details about the process used to curate this data. In summary, Buzzfeed curated the list using two journalists who were dedicated to identifying partisan news. Websites were added based on the criterion that ``they are clear about their political views and alignment and/or are explicitly ideological in their identity''.  Although it is hard to include every eligible website (for many reasons, e.g., new ones are constantly created, and others lose traffic, as discussed in \S\ref{sec:Owners}), an effort was made to ensure that the dataset is reasonably complete. After collecting a seed set of 500 sites, the journalists identified other links on the home pages of these publications to further expand the dataset. In addition, the authors of this paper leveraged Alexa traffic referral data (\ref{sec:our-alexa}), and compiled a list of 78 potential websites that are among the top 5 referers of traffic to the websites in the dataset. This list was passed to our collaborators in Buzzfeed News. After a manual examination by them, only 8 of the 78 were deemed as important (and added to the list of partisan Websites), which provides an indirect indication of the completeness of the original list. 

Altogether, this process identified 668 Websites which form our base dataset. Each website in the final list was classified by the Buzzfeed journalists as `left' or `right' based on self identification on the website or its associated page, or if an overt ideological preference could be deduced by manual examination. The rest of our analysis relies on the manual coding of sites into `left' and 'right' by the two professional journalists. While there are many more right-leaning sites than left-leaning ones, we believe that this is due to difference in hyper-partisan activity between the two sides of the political spectrum. Where relevant to the result, we take care to adjust for this imbalance in the dataset.
\subsection{Augmented Dataset}
We augmented the data provided by Buzzfeed News by collecting information from a diverse set of data sources that can shed light on the traffic and other engagement metrics of  the partisan websites and their Facebook pages. This included using Alexa Web Analytics, gathering engagement counts and other metrics from Facebook, examining short URLs to determine click through rates.

\subsubsection{Alexa}
\label{sec:our-alexa}
Alexa is an internet portal which provides web analytics including website traffic statistics, site comparisons, and website audience. Alexa's measurement panel is based on a diverse set of over 25,000 browser extensions and plug-ins used by millions of people~\cite{AlexaBlog}. Alexa does not reveal the full methodology it uses to collect data, but notes that the current implementation extends beyond the common understanding that it is based on a single popular browser toolbar~\cite{AlexaBlog}.

Although Alexa's data collection methodology may lead to some biases, it is believed to be the best available option for demographic and traffic information on the Web~\cite{vaughan2013web}. Alexa provides a basic notion of reliability by filtering data and only reporting statistics about a website when it believes that the website has sufficient measurable traffic available\footnote{https://support.alexa.com/hc/en-us/articles/200449744}. Our choice of using Alexa is based on these factors.

To gain further insights about the traffic flow patterns of the partisan sites, we crawled several  metrics from Alexa between June--July 2017 for the websites. The main metrics gathered were the current traffic rankings,  the variation in traffic rank for the past one year (this is available as a chart; we crawled and converted this into actual ranks), top-5 upstream sites (sites which the users visit before visiting the site in question), and audience demographics, such as the distributions of age ranges, ethnicity, gender, education and income levels of the visitors to each site. 

According to Alexa, the rank data is derived from the traffic data provided by the users in Alexa's Global Data Panel over a rolling 3 month period and is based on the browsing pattern of people in Alexa's Global Data Panel\footnote{https://support.alexa.com/hc/en-us/articles/200449744-How-are-Alexa-s-traffic-rankings-determined}.  Alexa says that the ``audience demographics data comes from voluntary demographics information submitted by people in our global traffic panel. The data is for the past 12 months, updated monthly.''. Table~\ref{tab:alexa_dataset} provides details about the number of sites for which Alexa was able to provide statistically relevant numbers for each metric.   

\begin{table}[htb!]
\begin{center}
\footnotesize
\begin{tabu} to 0.35\textwidth {  X[l] |  X[r] }
 \toprule
\textbf{Alexa Metric} & \textbf{Size (no. of sites)} \\
\midrule
Top-5 upstream sites & 484 \\
 Ranks & 570 \\
 Rank charts & 586 \\
 Audience Demographics & 421 \\
\bottomrule
\end{tabu}
\caption{Statistics of the collected Alexa dataset}
\vspace{-0.8cm}
\label{tab:alexa_dataset}
\end{center}
\end{table}


\subsubsection{Facebook}
We augmented the Facebook dataset by collecting page metrics for each of the $507$ Facebook pages using Facebook's Graph API. The following metrics were collected
\emph{category}, \emph{fan\_count}, \emph{start\_type}, \emph{start\_date}, \emph{pages that follow these pages} of the pages, \emph{websites} associated with the pages and their \emph{verification\_status}. We also collected random samples of 1000 active \emph{user followers} for top 100 most popular partisan pages.

\subsubsection{Click counts from Short URLs}
Through some initial exploration of the Buzzfeed Facebook dataset we discovered that about 6\% of the 4M Facebook posts were short URLs that mostly linked back to the article on the website. In order to study the relationship between their click through rates and Facebook post engagement, we collected the cumulative clicks count data for a sample comprising of {25000} Bit.ly short URLs across 216 pages using the Bitly API.

\section{Producers of partisan news}
\label{sec:Owners}
We first look at the websites, by examining domain registrations. We look at who the owners are and when the websites were registered, relative to the US elections. We find several owners who run groups (i.e., more than one) of partisan websites. We also identify a number of websites which appear to have been created specifically for the US elections, registered in the run up to the elections, and then losing popularity immediately after.  

\mypara{Who are the owners}
We begin by asking who are the owners of these websites, using the ownership records of the website domain names as returned by WHOIS\footnote{The WHOIS protocol, as specified in RFC 3912 is a way to look up owners of Web domain names. The Internet Corporation for Assigned Names and Numbers (ICANN) provides a WHOIS primer with more details at https://whois.icann.org/en/about-whois} queries. 
We find that 78 of the 668 partisan sites  (i.e., 11\%) in our list are run by Macedonian owners, further confirming the story of the macedonian teens and young men who ran successful partisan websites~\cite{wired}. These are mostly conservative sites (74 of these 78 are right-leaning, 4 are left-leaning).
Additionally, we discovered that at least $12$ U.S. based firms manage multiple websites and associated Facebook pages as listed in table \ref{tab:top_owners}. 

\begin{table}[htb!]
\begin{center}
\footnotesize
\begin{tabu} to 0.45\textwidth {  X[l] | X[l]  }
 \toprule
 \textbf{Owner Companies} & \textbf{No. of websites owned}  \\
 \midrule
Discount Book Distributors & 22 right \\
Addicting Info & 8 left \\
DB Capitol Strategies & 6 right \\
Today's Growth Consultant & 5 right \& 1 left \\
Salem Media Group &  5 right \\
Power Publisher & 4 right \\
Media Research Center & 4 right \\
Liftable Media & 3 right \\
News Corpse & 3 left \\
Counter Punch & 2 left \\
\bottomrule
\end{tabu}
\end{center}
\caption{Top 10 companies by no.\ of partisan sites owned}
\label{tab:top_owners}
\vspace{-0.8cm}
\end{table}

Largely, we observe that most of the owner companies have overt political affiliations i.e. they either handle collections of conservative or liberal sites. One prominent exception to this is \emph{Today's Growth Consultant} that runs 5 conservative sites including \emph{My Right America} and \emph{Red White and Right} along with \emph{Progressive Liberal} which is a left leaning site. Such bipartisan ownership of partisan websites suggests that the motive behind running collections of partisan sites cannot solely be attributed to propaganda but might also include financial incentives (e.g., advertising revenue).

\begin{figure}[!htb]
\centering
\includegraphics[width=0.8\columnwidth]{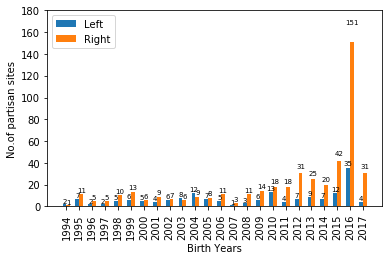}
\caption{Domain registration dates of partisan websites}
\label{fig:domain_registrations}
\end{figure}

\mypara{Birth and death of partisan websites}
Next, we analyse the role played by the US election timeline on the production of partisan content. We map the birth of these websites by using the date of the domain name registrations (Figure.~\ref{fig:domain_registrations}). 188 of the 668 websites  -- nearly a third -- were registered \emph{in the election year of 2016}, making it the year with the most number of domain name registrations for the sites in our dataset. 

Among these 188 websites, 81\% support conservative views while 19\% are liberal, pointing to an increased activity of partisan news on the conservative side. Note that this  activity is disproportionately large considering that about 73\% of the sites in our overall dataset are conservative.

Then we study the `death' or loss of traffic for these websites. We do this by acquiring the longitudinal Alexa rank numbers for the websites spanning across the period of one year.



\begin{figure}[!htb]
\centering
\includegraphics[width=0.8\columnwidth]{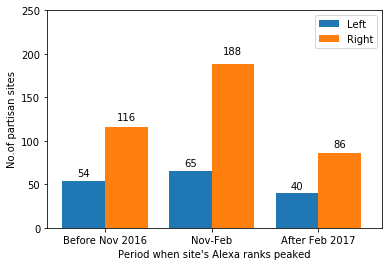}
\caption{Point of decline in popularity of partisan websites. The distribution of the date of peak popularity of individual websites is shown. From this point on, the traffic rank of the sites have been in overall decline (measured until June 29, 2017)}
\label{fig:temporal_popularity}
\end{figure}

Figure \ref{fig:temporal_popularity} provides the distribution of the time of peak Alexa rank (or equivalently, peak traffic) for the  partisan sites. About 31\% of the left-leaning and 40\% of the right-leaning sites had highest individual Alexa Ranks between Nov 2016 (when the election took place) to Feb 2017 (date when Trump moved into the White House). Again, we note that a disproportionately large number of conservative websites peaked and declined in popularity after this period. 

These results indicate that the traffic to these hyper-partisan sites is co-incident with the US elections. There could be several possible explanations. For instance, it may be that some websites were run by foreign actors or other agents with an explicit interest in promoting partisan news during the US elections, but stopped promoting their sites after the elections as it was no longer necessary. A second possibility is that  the website owners were financially motivated and the sites were no longer as valuable after the elections. Alternately, the consumers of this content were more interested during the closely fought elections and lost interest afterwards. Our data can only observe the correlation of the website registrations and peak popularity with the election cycle, and cannot distinguish between these different possibilities.




\section{Consumers of partisan news}
\label{sec:Alexa}

\begin{figure*}[!t]
\centering
\subfloat[]{
\includegraphics[width=0.22\textwidth, height = 4cm ]{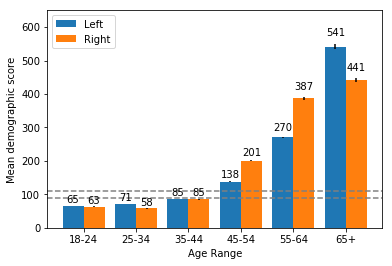}
\label{fig:Age}
}
\subfloat[]{
\includegraphics[width=0.22\linewidth, height = 4cm ]{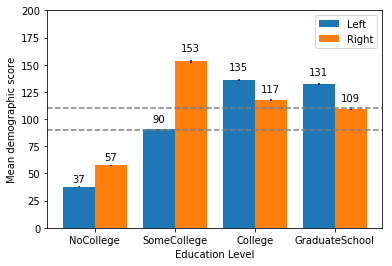}
\label{fig:Education}
}
\subfloat[]{
\includegraphics[width=0.22\linewidth, height = 4cm ]{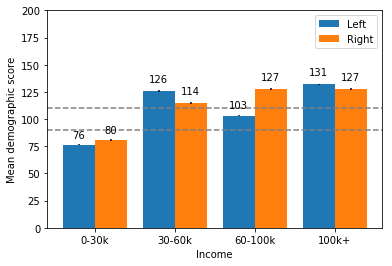}
\label{fig:Income}
}
\subfloat[]{
\includegraphics[width=0.22\linewidth, height = 4cm ]{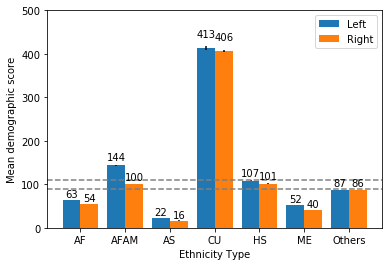}
\label{fig:Ethnicity}
}
 \vspace{-0.4cm}
\caption{ Demographic insights into audience of partisan websites, shown as  bar plots with standard error. X-axis shows  demographic bracket; Y axis shows sampled mean demographic score, using the scale in Table~\ref{tab:demo_scale}. Horizontal bands are drawn to indicate demographic scores in the range 90--110, which is considered to be representative of the general Internet population (Table~\ref{tab:demo_scale}). \ref{fig:Age}:  Age  Fig.\ref{fig:Education}:  Education ,\ref{fig:Income}: Income. \ref{fig:Ethnicity}: ethnicity. The ethnicities are coded as AF: African, AF/AM: African-American, AS: Asian, CU: Caucasian, HS: Hispanic, ME: Middle Eastern and finally Others.}
\vspace{-0.4cm}
\end{figure*}

Having considered the producers, we next seek to understand who are the consumers. To this end, we analyzed demographics data for partisan websites collected using the Alexa API, which also provides information about how the reported statistics compare with the general Internet population.
We were able to collect demographics data for 63\% (421/668) of the  partisan websites\footnote{Alexa does not have enough data about the remaining 247 sites. This may partly be explainable by the fact that 114 of the 247 sites were only launched in 2016/17.}, of which 69\% (290/421) are right-leaning and the remaining 31\% (131/421) are left-leaning.  Because of the differences in the numbers between the left and the right, we take  demographic scores for the left and the right by sampling equal numbers of sites, and report the mean of this sample.


\begin{table}[ht]
\begin{center}
\footnotesize
\begin{tabu} to 0.4\textwidth {  X[r] | X[l]  }
 \toprule
 \textbf{Range} & \textbf{Audience Representation} \\
\midrule
 0--50  & Greatly under-represented  \\
50--90 & Under-represented  \\
90--110 & Similar to the general Internet population   \\
110--200 & Over-represented   \\
200--Infinity & Greatly over-represented \\
\bottomrule
\end{tabu}
\caption{Audience Demographics Scale (from Alexa.com)}
\label{tab:demo_scale}
\vspace{-0.7cm}
\end{center}
\end{table}
We focused our analysis on five aspects of demographics namely age, education, ethnicity, income and gender. 
We visualize these patterns through a series of bar plots. The X axis represents the demographic bracket and the Y axis the sampled mean audience count (normalized to 100 for the general Internet population, as shown in Table~\ref{tab:demo_scale}) across all the left- or right- leaning websites. All bar plots are shown with standard error.

As expected from other surveys~\cite{pewSurveyNews4} that suggests younger people are less politically engaged, Figure~\ref{fig:Age} shows that those aged 45 and younger are under-represented relative to the rest of the Internet population, and those above 45 over-represented.  Interestingly, the under-represented populations are nearly equally left- and right-leaning, but the difference between left and right also grows with age. Partisan news consumers amongst the young (25 to 34) and the oldest (65+) age groups tend to be more liberal, while the middle-aged and older consumers (45 to 64) are more conservatively inclined. These findings are generally in line with conventional wisdom~\cite{hollandgeneration} and think tanks according to which older people tend to be more conservative than their younger counterparts.

However, among the 65+ age group that show maximum engagement in partisan news, there are relatively more liberal than conservative consumers. 
This could potentially be explained by the ``tree ring'' hypothesis, which contends that reaching young adulthood during certain historical periods has an effect on lifelong attitudes. In particular, increased liberalism  among Americans who reached young adulthood (defined as age 16) in the 1960s (and are therefore currently 65+) has been confirmed with evidence from the General Social Survey~\cite{longterm-survey}.


Figure~\ref{fig:Education} shows that people with no college education   do not engage much with partisan news. Those with the most common educational attainment level (some college) have significantly higher affinity with right-leaning sites than left-leaning ones. On the other hand, people with a higher educational background (those who have completed a college or graduate degree) are active consumers of partisan websites, and tend to be left-inclined. We also notice that people with very high levels of education (Graduate School) are less represented than those with some college~\cite{leighley1999race}.

As expected from the SocioEconomic Status theory~\cite{verba1987participation}, Figure~\ref{fig:Income} indicates that people from economically weaker backgrounds do not tend to participate politcally and are overall less represented (i.e., have lower demographic scores) on partisan websites. We observe that people in the \$30-60k and \$100k+ income brackets are over-represented with nearly equal distributions of left-leaning and right-leaning readership. However, the upper middle-class audience (with income in the range \$60-100k) is more inclined towards conservative news. This is the most common income range in the USA and includes roughly 1 in 3 households in the USA\footnote{http://www.npr.org/sections/money/2012/07/16/156688596/what-americans-earn}.

Figure~\ref{fig:Ethnicity} shows the racial distribution of consumers of partisan websites. Caucasians represent an overwhelming majority of consumers as expected. Caucasian, Hispanic and Other audience is somewhat equally divided between left-wing and right-wing. African-Americans are found to be more left-leaning. We note that African, Asian, and Middle Eastern communities are under-represented for partisan websites in comparison with the rest of the Internet population; thus the audience for hyper partisan websites appears to be white caucasians.

We also note  (no figure included) that females are underrepresented amongst visitors to our list of partisan websites, and those that follow such websites are more left-leaning. We conjecture this is due to Hilary Clinton becoming the first female candidate to be nominated for president by a major U.S. political party (The Democrats). Male readers are over-represented for both conservative and liberal websites, but there is a higher inclination towards conservative news. The increased male support for conservative positions and candidates has been observed in several places (e.g., The Guardian \cite{guardianPost} 
and Washington Post \cite{washingtonPost}) and the results also resonate with  several surveys (e.g.,  Pew Internet Research~\cite{pewSurveyNews6}).

In summary, the demographics study identifies an increased representation of conservative-leaning audiences in the most common demographic categories for education, income and ethnicity. Furthermore, many of the more engaged (i.e., demographic score > 110) age ranges also tend to be right leaning. The only highly engaged demographic group which is left leaning is the 65+ age range ($\approx$ 15\% of the US population). 

\section{Partisan Traffic Flows} 
\label{sec:TrafficFlows}
Since the Random Surfer model was introduced in the  PageRank~\cite{page1999pagerank} paper, it has been understood that sites can become important if other important sites refer to them. In this section, we use this concept to understand how the partisan websites gather traffic. Using Alexa data, we obtain and analyse the list of upstream referers\footnote{We use the spelling `referers' [sic] following the original misspelling in the HTTP Referer header~\cite[pg 524]{httpbook}.} to the partisan websites in our dataset, finding strong evidence of partisan echo chambers of right- and left-leaning websites. 

\subsection{Drivers of traffic to partisan websites}
\label{referrerCategories}
 According to Alexa, upstream sites are sites that the visitors of a website visit immediately before visiting it. These are obtained using a diverse range of browser extensions and plugins from Alexa's measurement panel.  Depending on availability of statistics, Alexa lists up to 5 top upstream referers to each website. We use Alexa's list to understand who drives traffic to the partisan sites dataset. Upstream data is available for 484  of the 668 partisan websites. Of these, 70\% (339) are right leaning, and 30\% (145) are left leaning. 

As a baseline, we also examine the distribution of upstream traffic referers for a list of top 500 news websites ranked by Alexa\footnote{https://www.alexa.com/topsites/category/News}. These are mainstream media sites -- large news conglomerates that influence a vast readership and whose political news coverage is not \textit{formally} announced to be in alignment with a specific political category or ideology, and include websites such as \textit{cnn.com}, \textit{nytimes.com}, \textit{theguardian.com}, \textit{washingtonpost.com}, etc. For each mainstream website, we  collected upstream site information from Alexa, and were able to get data for 478 websites.

\begin{table}[ht]
\begin{center}
\footnotesize
\begin{tabu} to 0.9\columnwidth {  X[l] | X[l] | X[l]  }
 \toprule
 \textbf{Category} & \textbf{Description} & \textbf{Examples} \\
 \toprule
 Search  & Search engines  & Google, Yahoo, Bing  \\
Social & Social media sites & Facebook, Twitter \\
YouTube & Video-streaming  & YouTube \\
Internal & Partisan sites & Infowars, Breitbart  \\
External & Other sites & New York Times, Huffington Post \\
\bottomrule
\end{tabu}
\caption{Categories of upstream sites}
\label{tab:referer_categories}
\vspace{-0.8cm}
\end{center}
\end{table}
We divide the upstream referers into five broad categories as described in Table~\ref{tab:referer_categories}. As with many other websites, search engines and social networks such as Facebook are the main referers, apart from YouTube, which contributes a small but significant minority. The remaining two categories are `internal' and `external'. We define internal upstream referers as other sites within the same category. Thus, the internal referers for partisan websites are other partisan websites; internal referers for mainstream news websites are other mainstream news websites. The external category captures all sites which are not otherwise categorised (i.e., all non-internal, non-social and non search-engine and non YouTube websites).

\begin{figure}[!tb]
\centering
\vspace{-0.2cm}
\includegraphics[width=0.8\columnwidth]{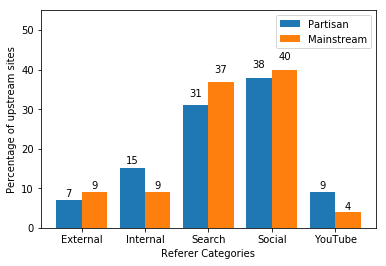}
\caption{Distribution of  Partisan and Mainstream referers }
\label{fig:Distribution of referers}
\vspace{-0.5cm}
\end{figure}

Figure~\ref{fig:Distribution of referers} shows that social networks and search engines are the top drivers of traffic to both partisan as well as mainstream news websites. However, partisan websites have a disproportionately large amount of internal referrals. The Internal category consisting of 14.7\% of partisan upstream sites that drive traffic among themselves together form the 3rd largest source of traffic to partisan sites as shown in Figure \ref{fig:Distribution of referers}. The share of traffic from the internal linkage between partisan websites is higher than many traditional drivers of internet traffic such as Twitter (9.3\%)  and  YouTube (9\%). We find this trend to be consistent for both left- and right-leaning websites (right-left split not shown in figure). 

Moreover, the share of internal upstream referers is 1.6 times higher in partisan websites than mainstream news sites. In terms of  individual numbers, nearly half (46\%) of the unique upstream sites to partisan websites are internal, in contrast to the case of mainstream websites where only 23\% of upstream sites are internal. These evidences point to an echo chamber of sorts amongst partisan websites which are driving traffic amongst each other, which we investigate further.

\subsection{Echo chambers among partisan producers}

\label{referrerBreakdown}

We previously observed in Figure~\ref{fig:Distribution of referers} that internal referers are a significant driver of traffic to partisan websites. We further break down such referring by political inclinations of the partisan websites to illuminate traffic flows among left and right websites.  
To test if we  see any peculiar internal linking among left- and right-leaning websites we plotted  the instances where a left-leaning partisan site links to a left-leaning site, and right-leaning one links to another right-leaning site. We also look at  sites that link across the partisaan divide (i.e., a right-leaning site linking to a left-leaning website and vice versa). 

\begin{figure}[!htb]
\centering
\vspace{-0.2cm}
\includegraphics[width=0.8\columnwidth]{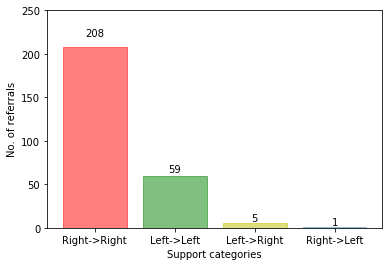}
\caption{Internal referrals within and across partisan divide}
\label{fig:internal_referring}
\vspace{-0.4cm}
\end{figure}

Figure~\ref{fig:internal_referring} shows that internal referrals are clearly split along party lines---there is a very significant amount of internal referrals within the right as well as within the left but very little referrals from left to right or right to left. Indeed, there are 208 instances of a right
partisan website being a top-5 referer to another right partisan
website, and 59 instances of a left partisan website being a top-5
referer to another left partisan website, in comparison to the single
digit number of instances where a left site is a top-5 referer for a
right-leaning site or vice versa. In total, partisan referrals between right-right or left-left account for 267/273 internal traffic referrals, i.e., $\approx$98\% of referrals between websites in our dataset happen within their own biased part of the political spectrum.  


Internal referrals appear to be more prevalent on the right: As mentioned in \S\ref{referrerCategories}, upstream traffic data is available for 339 (145) sites on the right (left). Of these, 139 (46) right (left) sites receive internal referrals according to Figure~\ref{fig:internal_referring}. In other words, 41\% of sites on the right benefit from receiving internal referrals as opposed to 31.7\% of sites on the left---internal referrals are 1.3 times more prevalent on the right than on the left.

This result highlights the existence of echo chambers among partisan websites in general, and conservative websites in particular. Such referral structures might serve to further reinforce and promote partisan micro-cultures.   

\begin{figure}[!ht]
\centering
\vspace{-0.4cm}
\includegraphics[width=0.8\columnwidth]{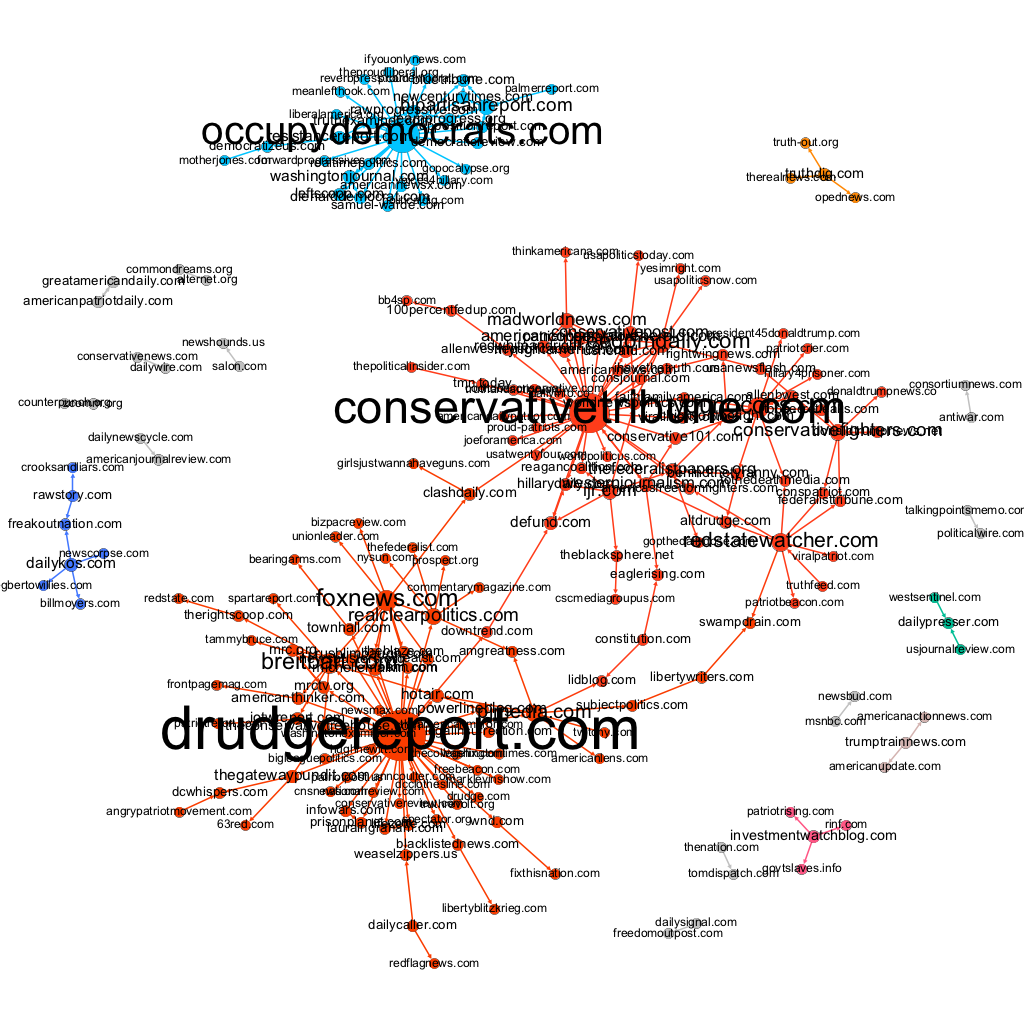}
\vspace{-0.4cm}
\caption{Partisan Sites Network Graph: Edges are drawn from site A to site B if A is one of the top 5 referers of traffic to B. In-degree is therefore capped at 5. Node size is proportional to out-degree, and large nodes are highly influential in directing traffic to other nodes.}
\vspace{-0.4cm}
\label{fig:partisan}
\end{figure}

\subsection{Network of partisan links}
To discover the most influential partisan domains we modelled the partisan sites as a network where an edge is drawn from one website A to another website B if A is a top-5 referer of B. Thus, we obtain a graph with a maximum in-degree of 5. We draw this graph in Figure~\ref{fig:partisan}, with right (left) leaning sites  coloured as red (blue), and the size of the labels is proportional to the out-degree of the node, i.e., the size of a node A is proportional to the number of other sites on whose top-5 referer lists A appears. This figure clearly visualises the separation between right and left leaning sites. This is confirmed by running the Louvain method for community detection, which achieves a modularity value of 0.69, indicative of an extremely strong community structure, with dense links within nodes in the same community and few links across communities.

We next move to identify the important players that send traffic to other sites, as the nodes with the largest number of outgoing edges, and the important beneficiaries as the nodes with the largest number of incoming internal edges. 
Table \ref{tab:hubs_and_authorities} lists the top 10 partisan sites with the largest outgoing and incoming links. Considering the ratio of right to left sites in our dataset (\S\ref{sec:Dataset}), we expect right-leaning sites to comprise $\approx$ 70\% of the lists. While we see the expected number of left-leaning sites amongst the sites with the most incoming links, the preponderance of right-leaning sites in the list of sites with the most outgoing links suggests the right has a monopoly on forward traffic referrals.


\begin{table}[ht]
\begin{center}
\vspace{-0.2cm}
\begin{tabu} to 0.4\textwidth {  X[l] | X[l]  }
 \toprule
 \textbf{Top out-degree} & \textbf{Top in-degree}  \\
 \midrule
 Drudge Report (R) &  Blue Tribune (L) \\
 Conservative Tribune (R) & Powerline Blog (R) \\
 Occupydemocrats (L) &  My Right America (R) \\
Fox News (R) & Raw Progressive (L) \\
Young Cons (R) & Hillary Daily (R)\\
Freedom Daily (R) & Hot Air (R) \\
Breitbart (R) & Die Hard Democrat (L) \\
Red State Watcher (R) &  Conservative101 (R)\\
Pj Media (R) & Red White and Right (R)\\
Conservative Fighters (R) & Rush Limbaugh (R)\\
\bottomrule
\end{tabu}
\caption{Top-10 websites by highest out-(in-) degree. Right-(Left-)leaning websites are marked as R (L). }
\vspace{-1.0cm}
\label{tab:hubs_and_authorities}
\end{center}
\end{table}

\section{Partisan Facebook Communities}
\label{sec:Facebook}
Many of the partisan websites in our dataset also have corresponding Facebook pages, with some of them having millions of followers. The social features of Facebook pages allows us to provide a more complete, yet different picture of the hyper-partisan news ecosystem: Rather than having to infer traffic referrals from Alexa summaries, we can identify which pages `like' which other page. Users who interact through comments and likes on Facebook pages can be identified, allowing us to directly examine the consumers of hyper partisan content. Short URLs such as bit.ly links are used on some Facebook pages. By tracking these links, we can identify engagement beyond comments and likes to measure click-through rates. Together, these studies paint a picture of the consumers of hyper partisan news as highly engaged but separated communities.

\subsection{Partisan page endorsements}
\begin{figure}[!ht]
\centering
\vspace{-0.4cm}
\includegraphics[width=0.8\columnwidth]{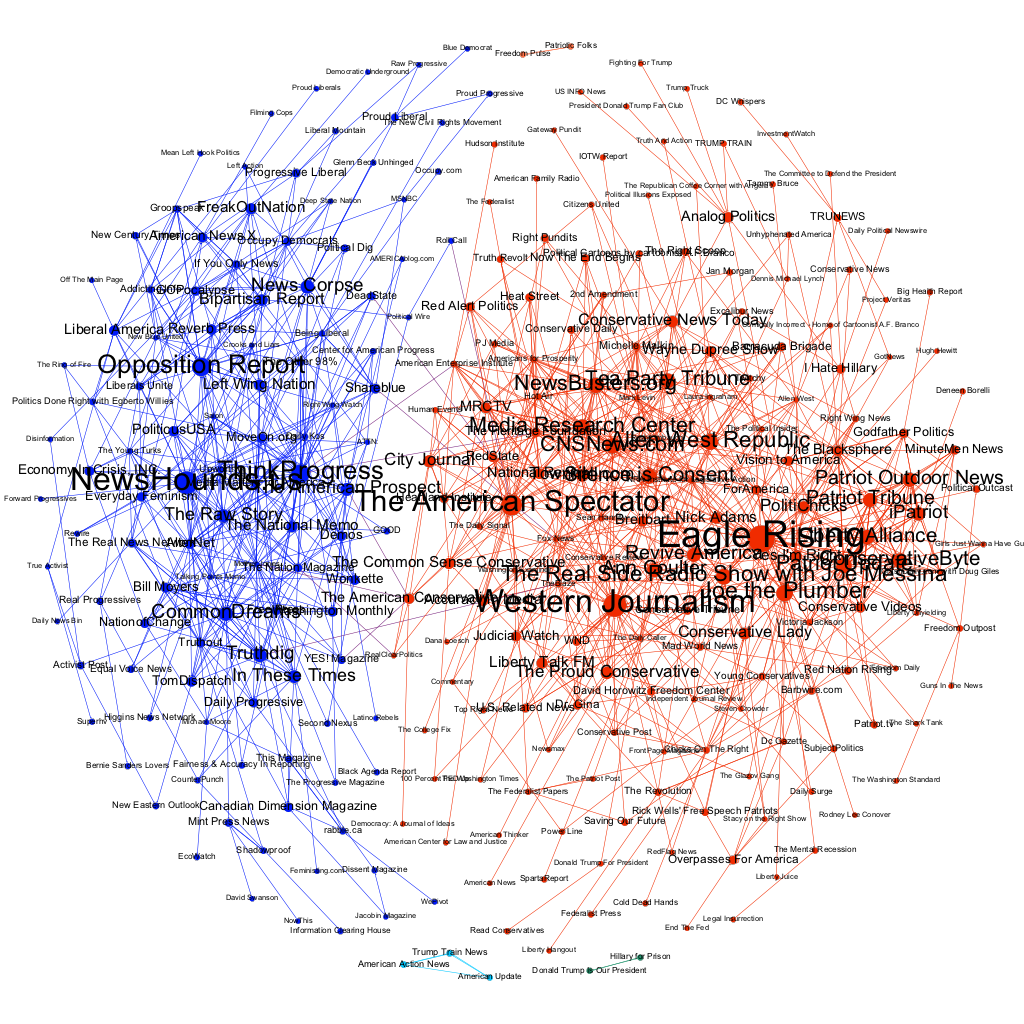}
\vspace{-0.2cm}
\caption{Graph induced from Facebook pages of partisan websites. Edges are drawn from page A to page B if  page A  `likes' B. Node sizes are proportional to the in-degree.}
\label{fig:facebookpages}
\vspace{-0.4cm}
\end{figure}
Our investigation of upstream sites and their internal traffic forwarding patterns showed a highly polarized structure of the partisan news web. However, this traffic referral had to be inferred through Alexa data, and the study was limited by the fact that only the top-5 referers were available through Alexa. By contrast, Facebook pages can explicitly `like' other pages, which allows us to infer the links between different partisan entities directly. 

Since many of the partisan websites we examine also have official Facebook pages, we crawl these using the Facebook Graph API, and also crawl the pages liked by these partisan pages. We then build a graph of relationships between these pages, assigning directed links from Page A to Page B if Page A likes Page B. Nodes are coloured red or blue according to whether they are right or left leaning. The resulting graph is shown in the Fig \ref{fig:facebookpages}. As with Fig.~\ref{fig:internal_referring}, this graph shows a clear community structure with a high modularity value of 0.46. The node labels are drawn proportional to the in-degree, i.e., the label of a page X is proportional to the number of pages which like page X. 

\subsection{Echo Chambers among Facebook Users}
To understand user engagement with these pages, we focus on the top 10 pages on the left and right with the most number of followers. In each page, we then identify 100 posts with the highest number of reactions. In this manner, we create a corpus of $10*100=1000$ posts on the left, and a similar number on the right. For each post, we identify up to 1000 users (API limit) who have liked the post. Thus, we identify up to $1000*100=100,000$ users who have liked posts on each Facebook page. 

With this data, we ask what is the overlap between the left and the right. We randomly pick $k=2$ pages on the left and right respectively. We then take the set of users who have interacted with the top 100 posts on the $k=2$ pages on the left and the set of users who have interacted with the top 100 posts on each of the $k=2$ pages on the right, and compute their intersection, finding the number of users who are common. We may similarly compute the left-left and right-right overlap, finding users who have interacted with $k=2$ pages on the left (right). This process can be repeated for $k=2,3,4\ldots 10$. Note that for any given $k$, the number of pages involved in a left-left or right-right interaction is $k$, whereas the number of pages involved in a left-right interaction is $2k$, thus offering \emph{more} opportunity for users to interact.

\begin{figure}[!h]
\centering
\includegraphics[width=0.8\columnwidth]{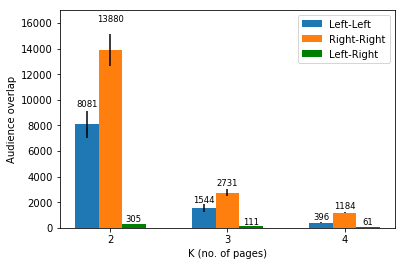}
\vspace{-0.4cm}
\caption{\textsl{\footnotesize Overlaps of users amongst K random pages from either sides }}
\label{fig:k_node_overlap}
\vspace{-0.6cm}
\end{figure}

Figure.~\ref{fig:k_node_overlap} shows the result of computing the overlap between pages using this process. As $k$ increases, the number of users who are common across all $k$ sites drops, and the values become too small to report after $k=4$. We also observe a left-left and right-right bias, wherein there is significantly more overlap of users who interact with pages all from the left or all from the right, than with pages from both the left and the right. While it is expected that the number of users who interact with $k$ different pages drops as $k$ increases, the significant difference between left-left and right-right interactions on the one hand and left-right interactions on the other hand again points to a hyper partisan \emph{audience}, similar to the hyper partisan links found between \emph{producers}.



\subsection{Characterising Facebook Page Engagement}

\begin{figure}[!h]
\centering
\includegraphics[width=0.8\columnwidth]{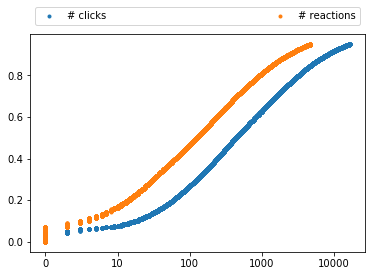}
\vspace{-0.3cm}
\caption{CDF of reactions on Facebook page and click through of bit.ly URL posts}
\label{fig:bitly_cdf}
\vspace{-0.35cm}
\end{figure}

The 4 Million posts that we obtained by crawling across 507 partisan pages, fall into three distinct categories. They are embedded URLs ($84.5$\%), photos ($11$\%) and videos ($3.7$\%). We also find that 6\% of the embedded URLs are shortened URLs, shortened by services like \texttt{bit.ly} or \texttt{goo.gl}. \emph{Tea Party}, \emph{ClashDaily.com with Doug Giles} and \emph{American Thinker} are the top 3 pages with maximum number of short-urls ($>15,000$ each). Finding \texttt{bit.ly} shortened URLs allowed us to track the click through rates of these shortened URLs using the \texttt{bit.ly} API.  In order to understand whether the Facebook users actually engage and if they do click through short URLs such as \texttt{bit.ly}, we crawled the clicks data for about 25 thousand short URLs belonging to 216 pages. The Fig. \ref{fig:bitly_cdf} shows the CDF of reactions on a facebook post with an embedded URL, and the click through for these URLs. Interestingly we note that click through is much higher than reactions on facebook. Also we found that the two variables are positively correlated (R = 0.26, p << $10^{-6}$). Moreover after normalising the clicks per subscriber by calculating $$mean(\frac{\textit{Clicks for a URL}}{\textit{Subscribers of the host page}})$$ across all partisan pages, we see a click through ratio of 4\%. This rate is, according to some sources\footnote{http://mashable.com/2009/07/07/twitter-clickthrough-rate/}, much higher than many marketing returns for ad campaigns. These insights show that Facebook pages are not only promoting the echo chambers of partisanship, but are also effective in accruing clicks for these sources.

\section{Conclusion}
\label{sec:Conclusion}
Hyper-partisan news media has been proven to be a major component in fake news proliferation\footnote{https://www.nytimes.com/2017/01/11/upshot/the-real-story-about-fake-news-is-partisanship.html}. In this study, we used a journalist-curated set of 668 hyper-partisan websites related to US News, and their associated Facebook pages, to characterise in depth the websites that produce hyper-partisan content, and the consumers of this content. We also answered two questions which have not been examined until now: how does traffic get directed towards partisan websites, and how do users engage with these websites. 

It was found that many of the websites, especially right-wing ones, were set up in 2016 during the run up to the US elections, and their Alexa traffic ranks declined after the Trump elections. This quick birth and death process  casts some doubt on the role of these sites as purveyors of `news' -- rather they appear to be makeshift platforms for distributing content when it was important to do so, or when there was a willing audience with an appetite for partisan politics. Our data-driven approach can only uncover the correlation of these websites' lifetimes with the US election. Pinning down the cause for this would be an  interesting avenue for future research.

We analysed the demographics of the audience of partisan websites, and observed an increased propensity for conservatism amongst the most populous demographic groups (e.g., Caucasians, middle income group, and average educational attainment), which points to increased interest in and potentially support for right wing causes. Analysis of traffic patterns over these partisan sites helped us uncover tightly interlinked communities of a partisan nature, which also showed evidence of echo chambers among producers of news, similar to ones observed amongst consumers previously. 

Overall, this study revealed that the way traffic, information and users are forwarded amongst hyperpartisan sites might be aiding, or worse, widening the highly polar nature of news media. Our study is not without limitations: As mentioned, a data-driven study such as this cannot uncover causal relationships. The dataset itself has been curated carefully, and to our knowledge, is the largest list of partisan websites. However, our conclusions may not generalise beyond those sites active in the 2016 US Elections, or to partisan sites in other geographies. Finally, although Alexa provides several hints and reassurances about the quality of data (see text and footnotes earlier), it does not fully reveal the methodology it uses, which makes it harder to understand the true potential or limitations of some of our results which rely on this data.

\balance
\bibliographystyle{ACM-Reference-Format}
\bibliography{bibliography} 

\end{document}